\documentstyle[12pt,psfig]{article}
\begin{document}
\begin{center}
{\bf Supercooling in viscous hydrodynamics for QCD phase transition}\\
\vspace{.2in}
P. Shukla, S. K. Gupta, A. K. Mohanty \\
{\it Nuclear Physics Division,Bhabha Atomic Research Centre,
Trombay, Mumbai 400 085, India}
\end{center}

\footnote{Ref: Phy. Rev. C {\bf 59}, 914 (1999)}
\begin{abstract}
\noindent
First order quark hadron phase transition is considered in scaling
hydrodynamics including the
Csernai-Kapusta model of nucleation of hadronic bubbles leading
to supercooling.
The effect of viscosity on
the entropy production is studied in ideal as well as in viscous
hydrodynamics with and without supercooling.
It is found that the excess entropy produced due to supercooling depends weakly  on
viscosity.
\end{abstract}

\section{ Introduction }
The ultrarelativistic heavy ion collisions provide a means to
create a new state of matter at high temperature and density
as the quantum chromodynamics (QCD) predicts a phase transition from
normal hadronic matter to quark gluon plasma (QGP), a state of
unconfined quarks and gluons. In recent years a considerable amount
of theoretical and experimental activity is going on in this field.
After the collision of two Lorentz contracted nuclei,
QGP is assumed to be formed and equilibrated in a very small time
of the order of $\sim$ 1 fm.
In the usual description of the evolution of the plasma, it cools by
expanding hydrodynamically till the critical temperature $T_c$ is
reached at which a transition from QGP to normal
hadronic matter takes place.
The temperature remains fixed at $T_c$ until
hadronization gets completed. This hadronization scenario corresponds
to the ideal Maxwell construction. However, in reality hadronization
does not begin at $T=T_c$ due to the large nucleation barrier.
Recently Csernai and Kapusta have proposed a model for nucleation for
the relativistic first order phase transition \cite{CSER}.
Supercooling, through nucleation of hadronic bubbles in QGP has been
studied by several authors \cite{CSER,CSER1,CSER2,DKS} using
the Csernai-Kapusta model of nucleation.
A general outcome of these studies is that the plasma will cool
according to the law $T(\tau)=T_0 (\tau_0/\tau)^{1/3}$ till $T_c$.
The matter continues to cool below $T_c$ until the temperature
goes down to about $\sim$ .8 $T_c$, where bubble formation and
growth becomes sufficient to reheat the system due to the release of
latent heat.
Compared to the idealized Maxwell construction the
supercooling delays the transition and leads to an extra entropy
production as the nucleation process allows dissipation around the
hadronic bubbles. The dynamical prefactor \cite{CSER} in the
nucleation rate includes quark viscosity
coefficients which bring dissipative effect in the medium. In fact,
the nucleation rate is limited by the ability of the dissipative
processes to carry latent heat away from the bubbles's surface.
However, for the dynamical evolution of the plasma the
ideal hydrodynamics is used.
This is not consistent as the viscosity dependent terms in
hydrodynamics would also contribute to the entropy production.
Therefore, in this work we study the supercooling in the viscous
hydrodynamics. The role of viscosity on supercooling and entropy production
has been investigated in detail.

\section{ Csernai-Kapusta model of nucleation }

 The nucleation model
computes the probability that a bubble of the
hadronic matter appears in a system, initially in QGP phase near the
critical temperature. Further in the model a small baryon chemical potential
is assumed. The bubble formation is assumed to take place
in a homogeneous QGP phase consisting of $u$ and $d$ quarks and gluons,
ignoring the role of inhomogenities such as strange quarks.
It is further assumed that there is not substantial supercooling.
Langer's theory of nucleation gives the nucleation rate per unit
volume at temperature $T$ as

\begin{eqnarray}\label{basic}
I=\frac{\kappa}{2\pi} \frac{\Omega_0}{V} e^{-\Delta F_*/T},
\end{eqnarray}
where $\Delta F_*$ is the change in the free energy of the system due to
the formation of a critical hadronic droplet, $\Omega_0$ is a
statistical prefactor
which measures the available phase volume and $V$ is the volume of the
system. The dynamical prefactor, $\kappa$ determines the
exponential growth rate of critical droplets which are perturbed from
their equilibrium radius $R_*$.
The coarse-grained effective field theory approximation to
QCD is utilised to obtain $\kappa$ and $\Omega_0$.
The factor $\Omega_0$ is obtained as

\begin{eqnarray}\label{omega}
\frac{\Omega_0}{V} = \frac{2}{3} \left(\frac{\sigma}{3T}\right)^{3/2}
\left(\frac{R_*}{\xi_q}\right)^4,
\end{eqnarray}
where $\sigma$ is the surface free energy and is determined as
50 MeV/fm$^2$ by lattice guage theory simulations without dynamical
quarks. The correlation length $\xi_q$ is estimated as 0.7 fm.
 The dynamical prefactor $\kappa$ is obtained as

\begin{eqnarray}\label{kappa}
\kappa = \frac{4 \sigma (4/3 \eta_q + \zeta_q) }
{(\Delta \omega)^2 R_*^3},
\end{eqnarray}
where $\Delta \omega$ is the difference in the enthalpy densities
of the two phases. $\eta$ and $\zeta$ are respectively, the shear and
bulk viscosity coefficients.
 Inserting $\Omega_0$ from Eq.~(\ref{omega}) and $\kappa$ from
Eq.~(\ref{kappa}) in Eq.~(\ref{basic}) we get the nucleation rate
per unit volume as

\begin{eqnarray}\label{rate}
I = \frac{4}{\pi} \left( \frac{\sigma}{3T} \right)^{3/2}
      \frac{\sigma (4\eta_q/3 + \zeta_q) R_*}{\xi_q^4 (\Delta \omega)^2}
     e^{-\Delta F_*/T}.
\end{eqnarray}

The critical radius $R_*$ is given by the Laplace formula as
\begin{eqnarray}\label{crit}
R_*(T) = \frac{2 \sigma}{p_h(T)-p_q(T)},
\end{eqnarray}
where $p_{q/h}$ is the pressure of the quark/hadron
phase at temperature $T$ and the $\Delta F_*$ as
\begin{eqnarray}\label{}
\Delta F_* = \frac{4\pi}{3} \sigma R_*^2.
\end{eqnarray}
  From the nucleation rate Eq.~(\ref{rate}), one can calculate the fraction
of volume $h(\tau)$ which has been converted from QCD plasma to hadronic
gas at proper time $\tau$.
If the system cools to $T_c$ at time $\tau_c$, then at some
later time $\tau$ the fraction $h$ of space which has been converted
to hadronic gas is

\begin{eqnarray}\label{frac}
h(\tau) = \int_{\tau_c}^\tau d\tau' I(T(\tau')) [1 - h(\tau')] V(\tau',\tau).
\end{eqnarray}
Here $V(\tau',\tau)$ is the volume of a bubble at time $\tau$ which had
been nucleated at an earlier time $\tau'$; this takes into account
the bubble growth. The factor $1 - h(\tau')$ accounts for the fact that
new bubbles can only be nucleated in the fraction of space not already
occupied by the hadronic gas. The model for bubble growth is
simply taken as \cite{WEIN}

\begin{eqnarray}\label{}
V(\tau',\tau) = \frac{4\pi}{3} \left( R_*(T(\tau')) +
\int_{\tau'}^\tau d\tau'' v(T(\tau'')) \right) ^3,
\end{eqnarray}
where $v(T)$ is the velocity of the bubble growth at temperature $T$.
By definition a critical size bubble is metastable and will not grow
without a perturbation. The growth of bubbles has been studied
numerically with relativistic hydrodynamics by Miller and Pantano
\cite{MILLER}. Their results are consistent with the growth law

\begin{eqnarray}\label{}
v(T) = 3 c [1 - T/T_c]^{3/2}.
\end{eqnarray}
This expression is intended to apply only when
$T > \frac{2}{3} T_c$ so that the growth velocity stays below the speed
of sound of a massless gas, $c/\sqrt {3}$.
At the critical temperature, $R_* \rightarrow \infty$,
 $\Delta F_* \rightarrow \infty$ and  the rate of
nucleation vanishes. The system must
supercool at least $\sim$ 5 \% to attain a finite rate.
 In the evolution of the matter from QGP to hadron phase, the temperature
varies with $\tau$ and its description by scaling hydrodynamics provides
another equation so that $h(\tau)$ and $T(\tau)$ are determined from
these equations at any proper time $\tau$.

\section{ Viscous hydrodynamics }

  The scaling viscous hydrodynamics is discussed by
Danielewicz and Gyulassy \cite{DAN} and others. The form of the
dissipative terms depends on the choice of the definition of
what constitutes the local rest frame of the fluid. The
Landau-Lifshitz definition is appropriate for describing systems
with small (or zero) chemical potential. Here we give a simple
derivation of the viscous hydrodynamics. In scaling hydrodynamics
the expansion takes place only along the direction of collision
which we chose as z axis.
The proper time $\tau$ and space-time rapidity $y$ are used in
place of $t$ and $z$ which are defined by

\begin{eqnarray}\label{}
\tau = \sqrt{t^2-z^2}
\end{eqnarray}
and
\begin{eqnarray}\label{}
y = \frac{1}{2} ln \frac{t+z}{t-z}.
\end{eqnarray}
 In the rest frame of the fluid we take a volume element as
$\delta V = A_{\bot} \tau \delta y$, where $A_{\bot}$
is the area of the element transverse to z direction.
The expansion takes place with the velocity $v_z = z/\tau$,
in the rest frame of the element.
Due to the viscous effects in longitudinal direction the heat
density per unit time is given by \cite{MIR}
\begin{eqnarray}\label{}
\phi & = & 2 \eta \left( \frac{\partial v_z}{\partial z} \right)^2 +
 (\zeta - 2 \eta/3) \left( \frac{\partial v_z}{\partial z} \right)^2, \nonumber \\
{\rm or} \hspace{1in} \phi & = & (4\eta/3 + \xi)/\tau^2.
\end{eqnarray}
After time increment $\Delta \tau$,
the volume expands by $\Delta V = A_{\bot} \Delta \tau \delta y$.
The amount of work done in expansion is $p \Delta V$.
 If the energy density at $\tau$ is $\epsilon$ and at $\tau+\Delta \tau$
is $\epsilon + \Delta \epsilon$ then the energy conservation implies
\begin{eqnarray}\label{}
\epsilon \delta V & = & (\epsilon + \Delta \epsilon) (\delta V + \Delta V)
          + p \Delta V - \phi \tau \Delta V, \nonumber \\
{\rm or} \hspace{1in} 0 & \simeq & \Delta \epsilon A_{\bot} \tau \delta y +
   (\epsilon + p - \phi \tau)A_{\bot} \Delta \tau \delta y,
\end{eqnarray}
 which leads to the scaling Navier-Stokes equation in the limit
$\Delta \tau \rightarrow 0$
\begin{eqnarray}\label{visc}
\frac{d\epsilon}{d\tau} &  = & -\frac{\epsilon + p}{\tau} +
        \frac{4\eta/3 + \zeta}{\tau^2}.
\end{eqnarray}
Equation~(\ref{visc}) has been
solved earlier \cite{JPG} for the QGP to hadron transition with the Maxwell
construction, i.e., assuming $T=T_c$ for the mixed phase.

To solve Eq.~(\ref{visc}) we require knowledge of the equation of state and
the temperature dependence of $\eta$ and $\zeta$.
In this work we use the bag equation of state for QGP. The energy density,
pressure and entropy densities in pure QGP and hadron phases are taken as
\begin{eqnarray}\label{}
\epsilon_q(T) = 3 a_q T^4 + B,\hspace{.1in}  p_q(T) =  a_q T^4 - B,
\hspace{.1in} s_q(T) = 4 a_q T^3,
\end{eqnarray}

\begin{eqnarray}\label{}
\epsilon_h(T) = 3 a_h T^4, \hspace{.1in} p_h(T) =  a_h T^4,
\hspace{.1in} s_h(T) = 4 a_h T^3.
\end{eqnarray}
Here, $a_q$ and $a_h$ are related to the degrees of freedom
operating in two phases and $B$ is the bag pressure.

  For ultrarelativistic gases, the bulk viscosity $\zeta$ is usually much
smaller than the shear viscosity $\eta$ \cite{WEIN}.
Danielewicz and Gyulassy \cite{DAN} give the acceptable range of $\eta$ for
the applicability of the Navier-Stokes equation to the expansion of the
plasma as
\begin{eqnarray}\label{range}
2 T^3 \le \eta \le 3 T^3 (\tau T).
\end{eqnarray}
 We define $\mu = (4\eta/3 + \zeta)$ and assume the temperature
dependence of viscosity coefficient as

\begin{eqnarray}\label{}
\mu_q(T) = \mu_{q0} T^3,
\end{eqnarray}
with $q$ being replaced by $h$ for the hadronic phase.

Solving Eq.~(\ref{visc})
for temperature in pure quark phase, we get \cite{JPG}

\begin{eqnarray}\label{}
T = T_0 \left(\frac{\tau_0}{\tau}\right)^{1/3} +
 \frac{\mu_{q0}}{8 a_q \tau_0}
\left[\left(\frac{\tau_0}{\tau}\right)^{1/3}-\frac{\tau_0}{\tau}\right]
\end{eqnarray}
for QGP formed at time $\tau_0$ while in pure hadron phase

\begin{eqnarray}\label{}
T = T_h \left(\frac{\tau_h}{\tau}\right)^{1/3} +
    \frac{\mu_{h0}}{8 a_h \tau_h}
\left[\left(\frac{\tau_h}{\tau}\right)^{1/3}-\frac{\tau_h}{\tau}\right],
\end{eqnarray}
where $\tau_h$ is the time where hadronization gets completed.

The energy density in the mixed phase at a time $\tau$ can be
written in terms of hadronic fraction $h(\tau)$ as
\begin{eqnarray}\label{}
\epsilon(\tau) & = &  \epsilon_q(T) + (\epsilon_h(T)-\epsilon_q(T))
       h(\tau), \nonumber \\
        & = & 3 [ a_q + (a_h-a_q) h(\tau) ] T^4 + B [1 - h(\tau)],
\end{eqnarray}
while the enthalpy density is
\begin{eqnarray}\label{}
\omega(\tau)  = 4 [ a_q + (a_h-a_q) h(\tau) ] T^4.
\end{eqnarray}
The temperature $T$ can be deduced as
\begin{eqnarray}\label{temp}
T(\tau) = \left(\frac{\omega(\tau)}{4 [ a_q + (a_h-a_q) h(\tau) ] }\right)^{1/4}.
\end{eqnarray}
 The shear and bulk viscosities are also taken to be the functions of time
according to

\begin{eqnarray}\label{}
\mu(\tau) & = &  \mu_q(T) + (\mu_h(T)-\mu_q(T)) h(\tau), \nonumber \\
{\rm or} \hspace{1in} \mu(\tau) & = & [ \mu_{q0} + (\mu_{h0}-\mu_{q0})h(\tau)]
                           (T(\tau))^3.
\end{eqnarray}

For the Maxwell construction, i.e., for $T=T_c$ in the mixed phase,
solution of Eq.~(\ref{visc}) is

\begin{eqnarray}\label{consh}
h(\tau) = (c-ab)[ Ei(b/\tau) - Ei(b/\tau_c) ] e^{-b/\tau}/\tau
   + (a+1) [1- \tau_c/\tau e^{(b/\tau_c - b/\tau)}].
\end{eqnarray}
Here, $Ei$ is exponential integral and

\begin{eqnarray}\label{}
a = \frac{4}{3} \frac{e_h}{e_q-e_h},
\hspace{.1in} b = \frac{\mu_q-\mu_h}{e_q-e_h},
\hspace{.1in} c = \frac{\mu_h}{e_q-e_h}.
\end{eqnarray}
The constants  a, b, and c are evaluated for $T=T_c$.

The solution given by Eq.~(\ref{consh}) does not
account for supercooling.
Due to supercooling, i.e., $T \ne T_c$ in the mixed phase, the temperature
is not constant and depends on $\tau$. Equation~(\ref{visc})
then becomes

\begin{eqnarray}\label{etau}
\frac{d\epsilon}{d\tau} = -\frac{\omega}{\tau}
+ \frac{\mu_{q0} + (\mu_{h0} - \mu_{q0}) h(\tau)}
{[ 4 a_q + 4(a_h-a_q) h(\tau)]^{3/4}}~~~ \frac{\omega^{3/4}}{\tau^2}.
\end{eqnarray}

Equation~(\ref{etau}) and Eq.~(\ref{frac}) are coupled equations,
finally to be solved for $h(\tau)$ and $\epsilon(\tau)$ [ or $T(\tau)$]. Once we
get $h(\tau)$ and $T(\tau)$ we can calculate the entropy density.

Eq.~(\ref{etau}) can also be written in terms of entropy as
\begin{eqnarray}\label{stau}
\frac{ds}{d\tau} = -\frac{s}{\tau} + \left(\frac{p_h(T)-p_q(T)}{T}\right)
~ \frac{dh}{dt}
+ \frac{\mu_{q0} + (\mu_{h0} - \mu_{q0}) h(\tau)}
{[ 4 a_q + 4(a_h-a_q) h(\tau)]^{2/3}}~~~ \frac{s^{2/3}}{\tau^2}.
\end{eqnarray}
Without the second and third terms on the right hand side, this equation
describes the conservation of entropy. The second term is responsible for
the entropy production due to nucleation. For the Maxwell
construction, $p_h-p_q=0$ as $T=T_c$ and this term vanishes.
The third term leads to continuous entropy production due to dissipative
effects.
We discuss the solution of these equations in the next section.

\section{ Results and Discussion}

By solving together Eq.~(\ref{etau}) and Eq.~(\ref{frac}), we have studied the
plasma evolution and calculated $s\tau$---the entropy production as
a function of $\tau$.
The initial conditions are taken at $\tau=\tau_c$ as $T=T_c$,
$h=0$ and $\epsilon(\tau_c)=\epsilon_q(\tau_c)$.
Now, $h(\tau)$ is calculated using Eq.~(\ref{consh}) with step
$\Delta$ in $\tau$. With this value of $h$, Eq.~(\ref{etau}) is solved for
$\epsilon(\tau)$ [or $T(\tau)$]. Then Eq.~(\ref{frac}) is evaluated by the
trapezoidal rule, thereby yielding new value of $h$.
Using new value of $h$, we solve again Eq.~(\ref{etau}) to
improve the value of $\epsilon$. This is repeated till an accuracy of
$\sim 10^{-5}$ is obtained. Then we proceed to the next step $\tau_c+2 \Delta$.
We take
$a_q = 37 \pi^2/90$
and $a_h = 4.6 \pi^2/90$.
The value 4.6 is taken instead of 3 to account for $\rho$,
$\omega$ and $\eta$ mesons apart from pions.
We chose $T_c$ = 160 MeV in this work implying
$B^{1/4}$ = 219 MeV.
For the hadronic matter $\eta_{h0} = 1.5$, $\zeta_{h0} = 1$ while
for the quark matter $\zeta_{q0} = 0$ \cite{DAN,JPG,HOSO}.
The coeficient $\eta_{q0}$ has been varied from 2.5 to 20.
The nucleation involves the surface free energy, correlation length,
and the velocity of bubble growth, which have already been
described in Sec. II.
The volume of a fluid element is $A_{\bot} \tau \delta y$.
As $\tau \delta y$ does not change with $\tau$ in the scaling
hydrodynamics, $s\tau$ provides a measure of total entropy.
So as to understand the role of supercooling
and viscous heat generation, we have
compared various scenarious; ideal hydrodynamics (IHD), IHD with supercooling,
viscous hydrodynamics (VHD) and VHD with supercooling.
We assumed
that at initial time $\tau_i$ = 1 fm/$c$, the temperature $T_i$=268 MeV marked the
beginning of the evolution. This corresponds to energy density $\epsilon$=8.51
Gev/fm$^3$ and entropy  $s_i\tau_i$=40.65 fm$^{-2}$. These are
appropriate for the Relativistic Heavy Ion Collider (RHIC) energies.
For further
discussions, it is instructive to examine $\tau$ versus $T$ plot. Figure 1 shows
such a plot for VHD calculations with supercooling included. From a initial
point ($\tau_i,T_i$), one moves on to $(\tau_c,T_c)$ where critical temperature
is reached. However, the nucleation rate is zero at this point and system
does not hadronize. The system keeps cooling to $(\tau_m, T_m)$ where
significant nucleation rate and hadronization is reached. The system cannot cool
further as entropy cannot decrease according to the second law of
thermodynamics.
At this point, the hadron fraction is around 11-18 $\%$ for various cases.
As the temperature increases towards $T_c$, the hadron fraction $h$ increases
and slowly approaches a value of one at the point $(\tau_h, T_c)$.
After this,  temperature starts decreasing finally reaching the point $(\tau_f,T_f)$
at the freeze-out temperature $T_f$. This is the general character of $\tau-T$
curve. If one uses the Maxwell construction, one reaches directly from
($\tau_c, T_c)$ to $(\tau_h, T_c)$ without any change in the temperature.

In Figs. 2 and 3, we show the results for
$T$, $h$ and $s\tau$ as a function of $\tau$ for $\eta_{q0}$=2.5 and 5 . For IHD, the critical temperature
$T_c$ is reached earlier than VHD. Similarly with supercooling included,
$T_m$ is
approached earlier in IHD than VHD. The supercooling leads to an abrupt
change in entropy starting from $\tau_m$ onwards, i.e., in the reheating
region.
We find in our calculations that $s\tau$ does not increase much beyond
$\tau \approx  $20 fm/$c$.
Therefore, $s_h\tau_h$ represents reasonably well the total entropy
produced. As $s_h$ is evaluated at $T_c$ for all cases, the total entropy
is also a measure of the life time of the system upto the hadronization
stage.
Figure 4 shows $s_h\tau_h$
as a function of $\eta_{q0}$. With VHD, entropy production increases
with viscosity. Supercooling leads almost to a constant shift of the curve for the VHD.
In Fig. 5, the excess entropy production due to supercooling is shown as a
function of viscosity. For all values of $\eta_{q0}$, excess entropy does not
change much. It is almost constant for the VHD calculation.
We also found that the values of $\eta_{q0}<<1$ do not lead
to significant hadronization and the system continues to cool in the QGP
phase. Sufficient value of $\eta_{q0}$ is required for the transition to the
hadron phase. The bubble growth velocity coefficient (3$c$ in our calculation)
also plays an important role in the transition as setting it to zero also
blocks the transition. The viscosity plays very crucial role in the transition
through nucleation. However, once viscosity enters into picture, the viscous
hydrodynamics is to be used for consistency. In the presence of viscosity, the initial
energy density estimates from the experimental rapidity density distribution
would be lower than the ones calculated with the Bjorken formula. Though
supercooling
does not lead to significant increase in entropy production, it is
very much needed. For the description of the phenomena in relativistic nuclear
collisions, the viscous hydrodynamics with supercooling is a more appropriate
framework than the ideal hydrodynamics with or without supercooling. Further
theoretical work is needed to compute the viscosity with narrower bounds for
the analysis of the experimental data.

\section{ Conclusions}
We have studied the entropy production both in
ideal and viscous scaling hydrodynamics with and without supercooling.
Excess entropy  produced due to
supercooling in viscous hydrodynamics, weakly depends on viscosity of the plasma phase.
Though, in general entropy produced due to supercooling is
much less than that due to viscous heat generation, the phenomenon
of supercooling provides an important physical mechanism
for the quark-hadron transition to occur. The viscous hydrodynamics with
supercooling also leads to an increase in the lifetime of the plasma.

\newpage
\begin{itemize}
\item FIG. 1.
$T/T_c$ as a function of proper time $\tau$ in viscous hydrodynamics
with supercooling.
\item FIG. 2.
The $T/T_c$, $h(\tau)$ and $s\tau$ as a function of $\tau$ in
IHD (dotted), IHD with supercooling (short dashed), VHD (long dashed)
and VHD with supercooling (solid) at $\eta_{q0}$=2.5.
\item FIG. 3.
Same as Fig. 2, but with $\eta_{q0}$=5.0.
\item FIG. 4.
$s_h\tau_h$ versus $\eta_{q0}$ in IHD, IHD with supercooling, VHD and VHD
with supercooling.
\item FIG. 5.
Excess entropy produced due to supercooling versus $\eta_{q0}$ both with
IHD and VHD.
\end{itemize}

\newpage
\begin{figure}
\begin{minipage}[t]{10cm}
\hspace{-8cm}
\psfig{figure=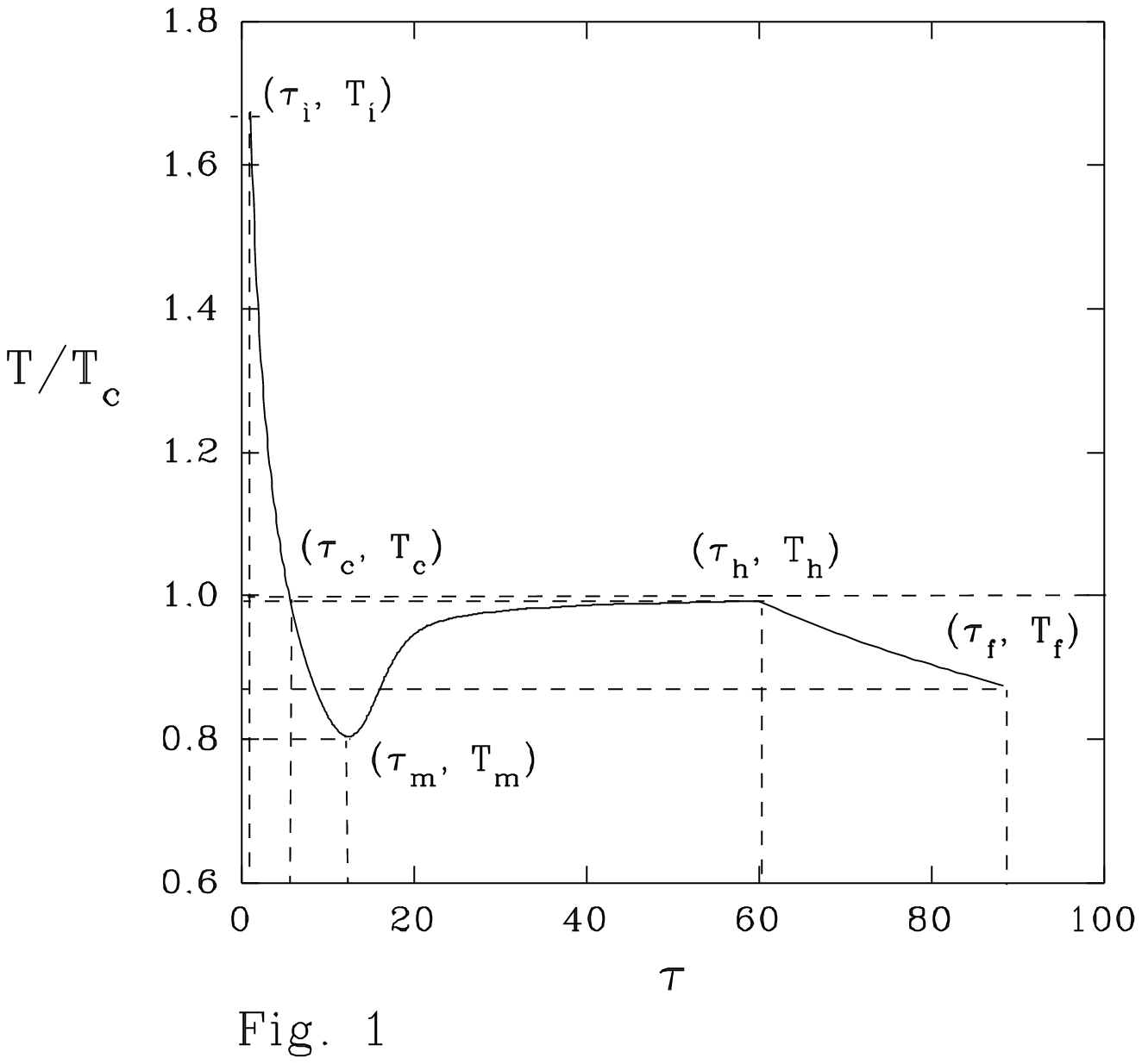,height=10cm, width=10cm}
\end{minipage}
\end{figure}

\newpage
\begin{figure}
\begin{minipage}[t]{10cm}
\hspace{-8cm}
\psfig{figure=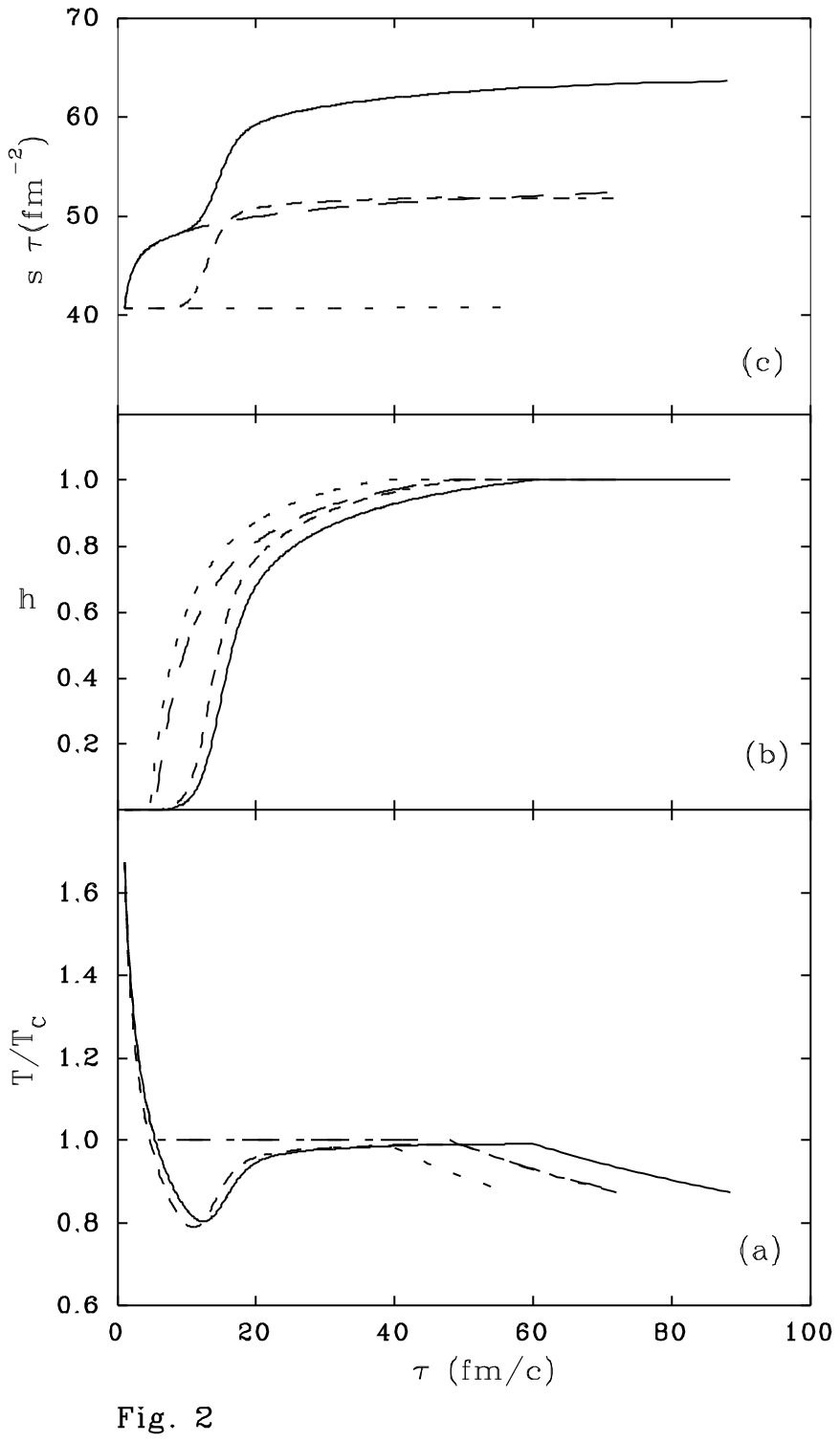,height=10cm, width=10cm}
\end{minipage}
\end{figure}

\newpage
\begin{figure}
\begin{minipage}[t]{10cm}
\hspace{-8cm}
\psfig{figure=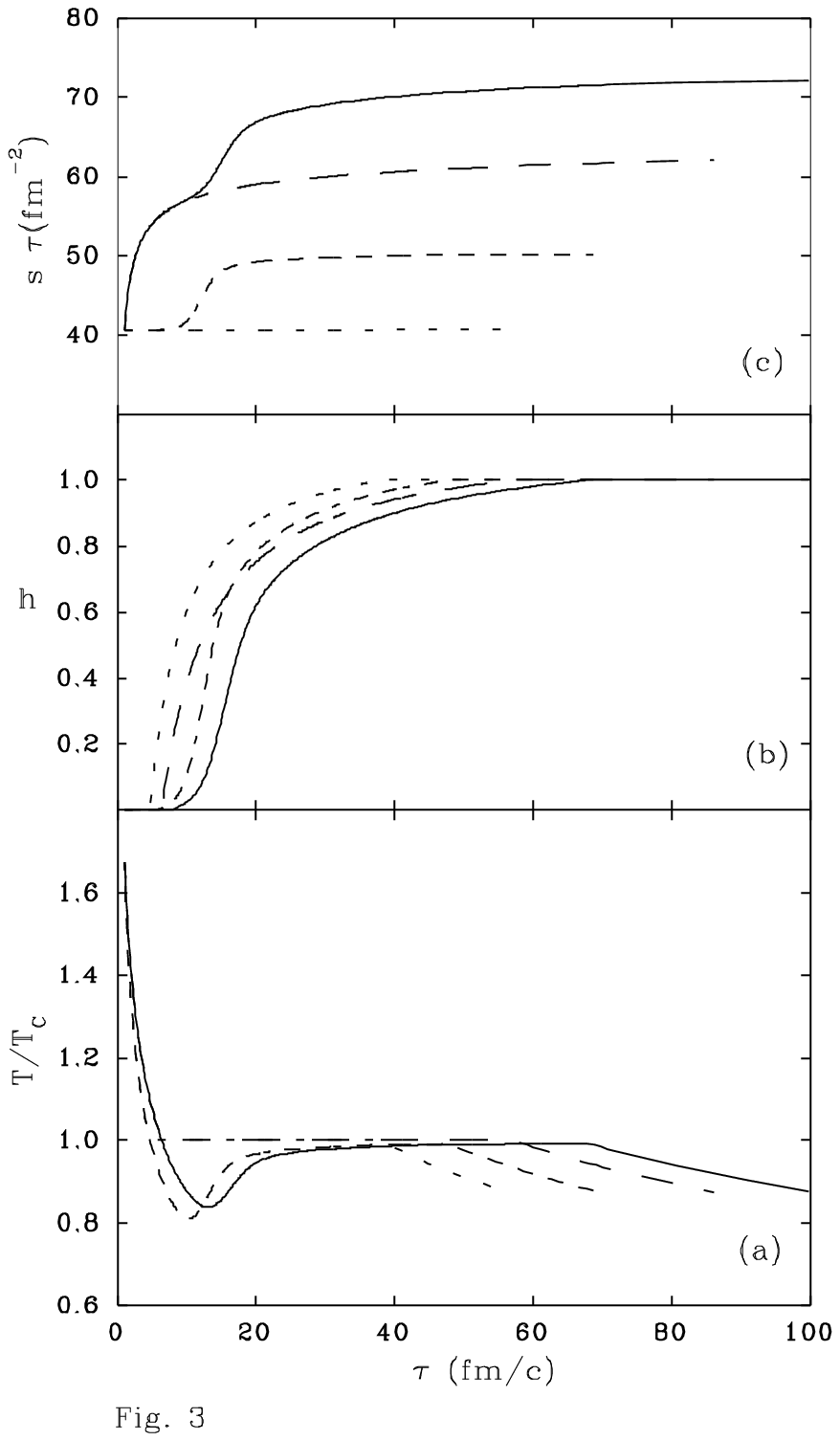,height=10cm, width=10cm}
\end{minipage}
\end{figure}

\newpage
\begin{figure}
\begin{minipage}[t]{10cm}
\hspace{-8cm}
\psfig{figure=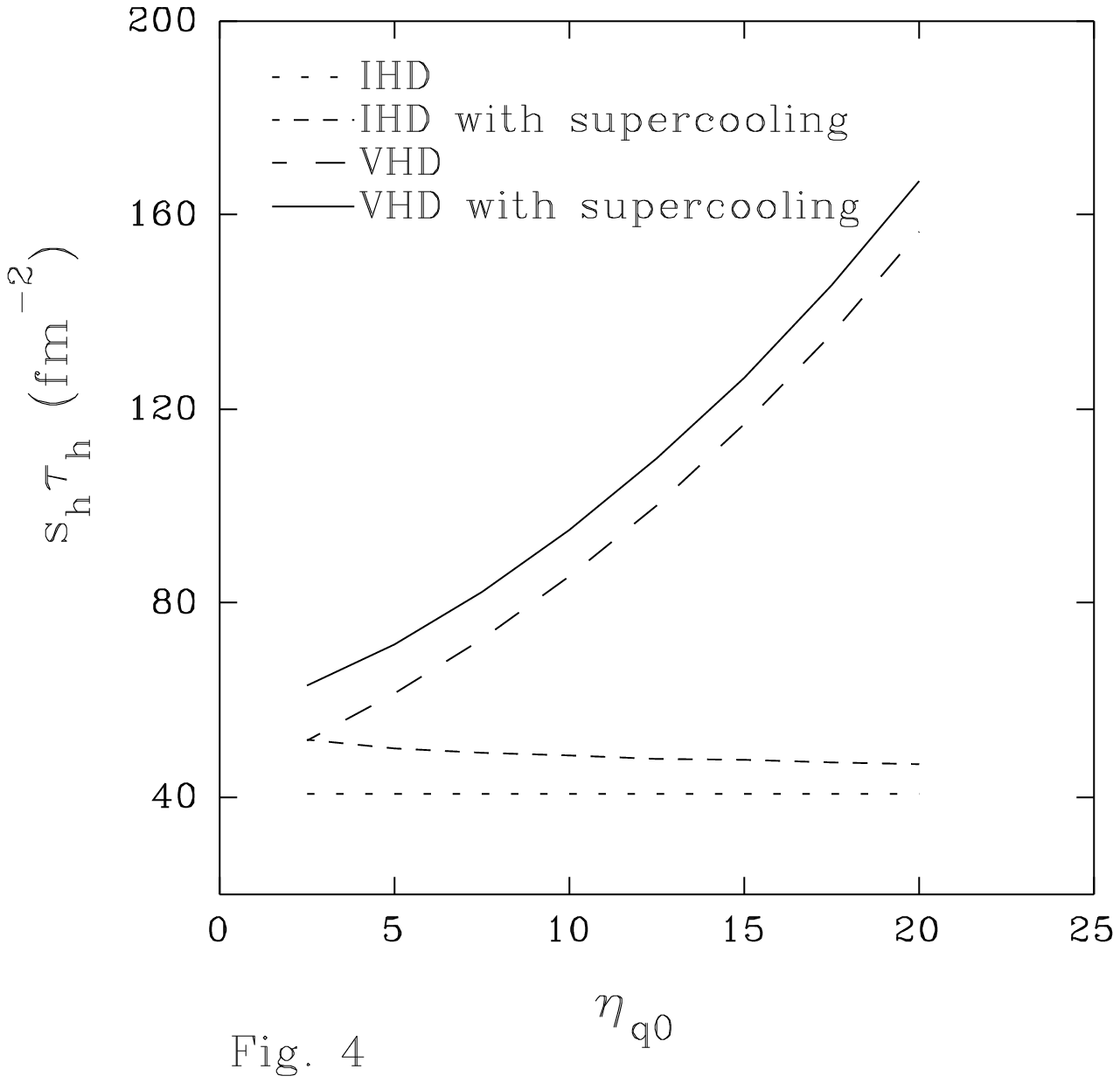,height=10cm, width=10cm}
\end{minipage}
\end{figure}

\newpage
\begin{figure}
\begin{minipage}[t]{10cm}
\hspace{-8cm}
\psfig{figure=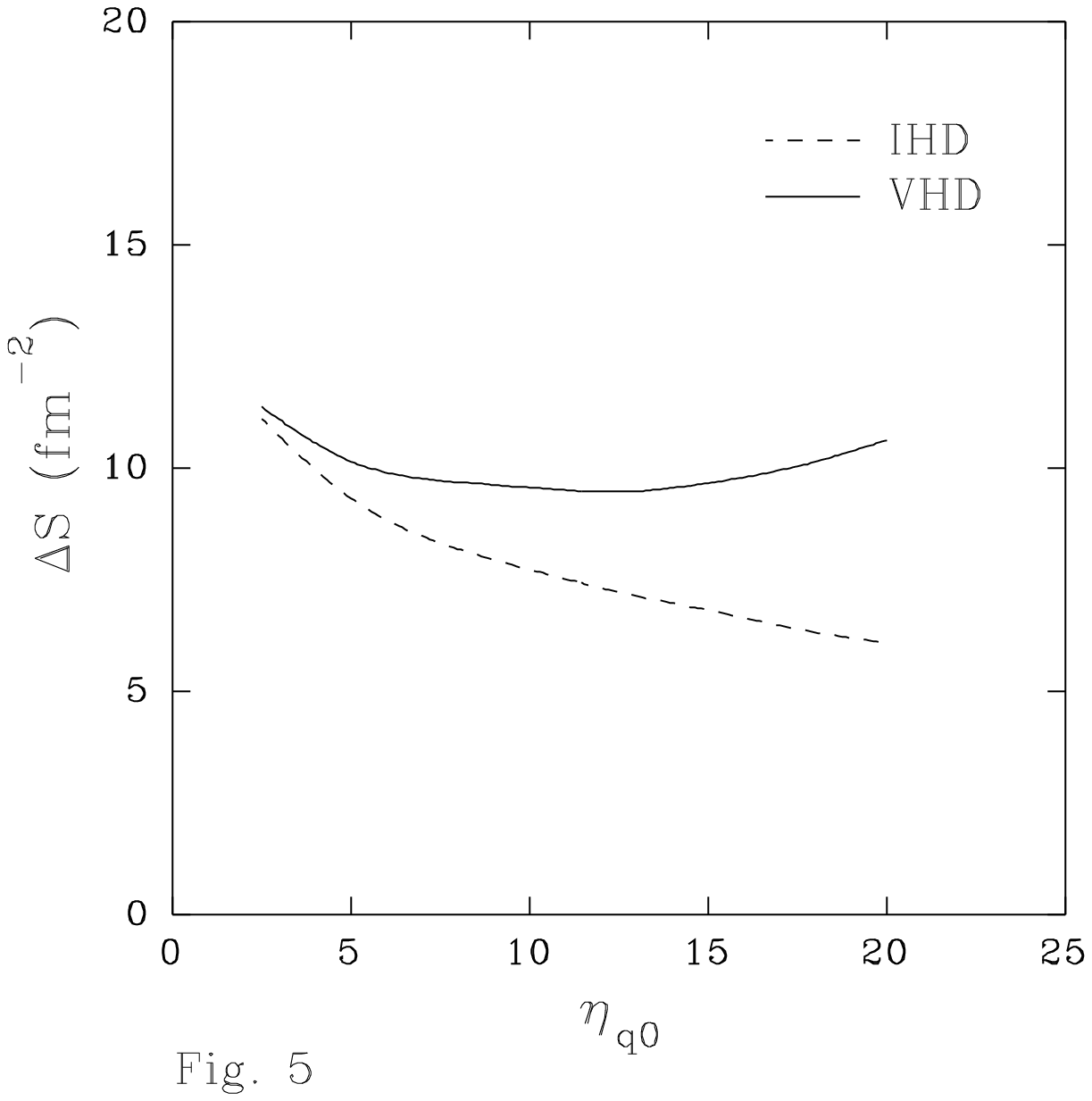,height=10cm, width=10cm}
\end{minipage}
\end{figure}


\begin{thebibliography}{99}

\bibitem{CSER} L. P. Csernai and J. I. Kapusta, Phys. Rev. Lett. {\bf 69},
 737 (1992); Phys. Rev. D {\bf 46}, 1379 (1992).

\bibitem{CSER1} L. P. Csernai, J. I. Kapusta, Gy. Kluge, and E. E. Zabrodin,
Z. Phys. C {\bf 58}, 453 (1993).

\bibitem{CSER2} T. Csorgo and L. P. Csernai, Phys. Lett. B {\bf 333}, 494 (1994).

\bibitem{DKS} M. G. Mustafa, D. K. Srivastava, and B. Sinha,
 nucl-th/9712014.

\bibitem{WEIN} S. Weinberg, Astrophys. J. {\bf 168,} 175 (1971).

\bibitem{MILLER} J. C. Miller and O. Pantano, Phys. Rev. D {\bf 40},
                 1789 (1989); {\bf 42}, 3334 (1990).
\bibitem{DAN} P. Danielewicz and M. Gyulassy, Phys. Rev. D {\bf31}, 53 (1985).

\bibitem{MIR} B. Yavorsky and A. Deltaf, {\it Handbook of Physics
        (English translation) } (Mir Publishers, Moscow, 1972), p. 333.

\bibitem{JPG} S. Sarkar, P. Roy, J. Alam, S. Raha, and B. Sinha,
       J. Phys. G {\bf 23}, 469 (1997).


\bibitem{HOSO} A. Hosoya and K. Kanjantie, Nucl. Phys. B {\bf 250}, 666 (1985).

\end{thebibliography}
\end{document}